\documentclass[%
 reprint,
 amsmath,amssymb,
prl,
]{revtex4-2}

\usepackage{lipsum}
\usepackage{bm}
\usepackage{amssymb}
\usepackage{amstext}
\usepackage{amsmath}
\usepackage{mathrsfs}
\usepackage{graphicx}
\usepackage{color}
\usepackage{amsthm}
\usepackage{floatrow}
\usepackage{array}

\usepackage{listings}
\usepackage{booktabs}
\usepackage{hyperref}

\begin{document}
\preprint{APS/123-QED}
\title{Laplacian dynamics of convergent and divergent swarm behaviors}
\thanks{Correspondence should be addressed to P.S.}%

 \author{Yang Tian}
\email{tiany20@mails.tsinghua.edu.cn}
 \altaffiliation[]{Department of Psychology \& Tsinghua Laboratory of Brain and Intelligence, Tsinghua University, Beijing, 100084, China.}
 \altaffiliation[Also at ]{Laboratory of Advanced Computing and Storage, Central Research Institute, 2012 Laboratories, Huawei Technologies Co. Ltd., Beijing, 100084, China.}

   \author{Yunhui Xu}%
\email{xuyunhui18@mails.tsinghua.edu.cn}
 \altaffiliation{Department of Physics, Tsinghua University, Beijing, 100084, China.}
 
\author{Pei Sun}%
 \email{peisun@tsinghua.edu.cn}
 \altaffiliation[]{Department of Psychology \& Tsinghua Laboratory of Brain and Intelligence, Tsinghua University, Beijing, 100084, China.}
 



\begin{abstract}
Swarming phenomena are ubiquitous in various physical, biological, and social systems, where simple local interactions between individual units lead to complex global patterns. A common feature of diverse swarming phenomena is that the units exhibit either convergent or divergent evolution in their behaviors, i.e., becoming increasingly similar or distinct, respectively. The associated dynamics changes across time, leading to complex consequences on a global scale. In this study, we propose a generalized Laplacian dynamics model to describe both convergent and divergent swarm behaviors, where the trends of convergence and divergence compete with each other and jointly determine the evolution of global patterns. We empirically observe non-trivial phase-transition-like phenomena between the convergent and divergent evolution phases, which are controlled by local interaction properties. We also propose a conjecture regarding the underlying phase transition mechanisms and outline the main theoretical difficulties for testing this conjecture. Overall, our framework may serve as a minimal model of swarm behaviors and their intricate phase transition dynamics.
\end{abstract}

\maketitle
\paragraph{Background.} Swarm behaviors evolve dynamically with time \cite{gazi2013lagrangian}. The convergent and divergent evolution processes of these behaviors exist across different scales. For example, in similar environments, multiple species that interact with each other may evolve along converging pathways \cite{khatri2009statistical,powell2015convergent,stayton2015does}, while the closely related populations of a species may evolve along diverging pathways if they experience distinct selective pressures \cite{sitarz2005divergent,vogwill2016divergent,zeldovich2008understanding}. As another example, people may exhibit convergent social behaviors (e.g., opinions or consuming behaviors) if emergencies occur (e.g., epidemic or social turmoil), while these social behaviors diverge and become increasingly individualized after emergencies end \cite{flierl1999individuals,castellano2009statistical,takayasu2015rumor}. 

Although the universality of convergent and divergent swarm behaviors ensures its ubiquity in the world, it does propose critical challenges for analytic modeling due to the rich connotations and diverse details of swarm behaviors. It is unclear whether the convergent and divergent dynamics of organic evolution, the convergent and divergent social behaviors of people, and any other similar phenomenon can be characterized by a unified and simple model. While specialized models \cite{tanaka2007general,hindes2016hybrid,piwowarczyk2019influence,czirok2001theory,yu2013swarming,wang2021emergent,yang2010swarm,thutupalli2011swarming,iwasa2012various,iwasa2017mechanism,huepe2008new}, statistical analyses \cite{attanasi2014finite,jose2022physical}, and data-driven methods \cite{xue2023machine} have achieved essential progress in studying swarming phenomena, advances in describing the convergent and divergent evolution processes of swarm behaviors remain limited. 

\paragraph{Our work.} In this study, we propose a simple model of convergent and divergent swarm behaviors and describe the dynamic switching of convergent and divergent evolution across time. This framework may serve as a minimal model to characterize diverse real swarming phenomena. In addition, our model can be enriched by considering more agent heterogeneity and local interaction rules (e.g., memory and preference). Our computational experiments of this model suggest phase-transition-like phenomena between convergent and divergent evolution phases, which are controlled by the balance between convergent and divergent tendencies in local interactions. We formalize a conjecture on the latent phase transition mechanisms and summarize theoretical difficulties for verifying this conjecture.

\paragraph{Local dynamics of convergent evolution.} To reach at a balance between the universality to fit in with different swarming phenomena and the simpleness to support analytic derivations, we consider a $n$-dimensional abstract state space $S\subset\mathbb{R}^{n}$, where each element $\mathbf{s}\in S$ denotes an abstract state that can be used to model any target properties of system units (e.g., spatial coordinates, opinions, or genetic characters). A swarm model updates the state of each unit $\sigma_{i}\in V$ at moment $t\in T$ according to its current state, denoted by $\mathbf{s}_{i}\left(t\right)=\mathbf{s}\left(\sigma_{i}\left(t\right)\right)$, and its interactions with its neighbor units, included by set $N_{i}$, following a local rule $\psi$
\begin{align}
    \mathbf{s}_{i}\left(t+\varepsilon\right)=\psi\left(\mathbf{s}_{i}\left(t\right),\mathbf{s}_{N_{i}}\left(t\right)\right).\label{EQ1}
\end{align}
Here $\mathbf{s}_{N_{i}}\left(t\right)=\{\mathbf{s}_{j}\left(t\right)\vert \sigma_{j}\in N_{i}\}$ stands for the states of neighbor units of unit $i$ at moment $t$ and $\varepsilon$ denotes a minimum time step (see Fig. 1(a) for illustrations).

The localization of interaction rules is realized by constraining $N_{i}$, the neighbor set of unit $\sigma_{i}$, as a sub-set of units whose states are close to $\sigma_{i}$
\begin{align}
    N_{i}=\{\sigma_{j}\vert d\left(\mathbf{s}_{i}\left(t\right),\mathbf{s}_{i}\left(t\right)\right)\leq\Delta\}\subset V,\label{EQ2}
\end{align}
where $d\left(\cdot,\cdot\right)$ denotes an arbitrary distance function and $\Delta$ is a parameter that determines the interaction range (see Fig. 1(a)). Please note that set $N_{i}$ evolves across time according to Eq. (\ref{EQ2}) as the unit state is time-dependent.

A convergent evolution process is defined as a global pattern of the system, during which units become increasingly similar to each other in the state space. This process also corresponds to the case where the states of all units approach the global mean state. The non-triviality of this process lies in that each unit only interacts with its neighbors and, consequently, can never know the absolute coordinates of all units in the state space. Only the relative coordinate information obtained through interactions is available for units at each moment. For each unit $\sigma_{i}$, the relative coordinate of each of its neighbors in the state space is defined as
\begin{align}
\mathbf{r}_{ij}\left(t\right)=\mathbf{s}_{j}\left(t\right)-\mathbf{s}_{i}\left(t\right).\label{EQ3}
\end{align}
Given this information, unit $\sigma_{i}$ can only approach the time-dependent local mean state
\begin{align}
\overline{\mathbf{s}}_{i}\left(t\right)=\mathbf{s}_{i}\left(t\right)+\frac{1}{\vert N_{i}\vert+1}\sum_{\sigma_{j}\in N_{i}}\mathbf{r}_{ij}\left(t\right)\label{EQ4}
\end{align}
during convergent evolution. Note that $\vert N_{i}\vert+1$ in the denominator of Eq. (\ref{EQ4}) is used to deal with the case with $\vert N_{i}\vert=0$.

The above definitions have described the convergent evolution process of swarm behaviors on a local scale. To study its non-trivial manifestation on a global scale, we need to relate Eq. (\ref{EQ2}) with specific global characteristics of the system. 

\paragraph{Global dynamics of convergent evolution.} We notice that Eq. (\ref{EQ2}) can be equivalently reformulated as
\begin{align}
\overline{\mathbf{s}}_{i}\left(t\right)=\mathbf{s}_{i}\left(t\right)+\frac{1}{\vert N_{i}\vert+1}\sum_{\sigma_{j}\in V}\mathbf{A}_{ij}\left(t\right)\left(\mathbf{s}_{j}\left(t\right)-\mathbf{s}_{i}\left(t\right)\right),\label{EQ5}
\end{align}
where $\mathbf{A}\left(t\right)$ denotes the time-dependent adjacency matrix
\begin{align}
\mathbf{A}_{ij}\left(t\right)=\left(1-\delta_{i,j}\right)\Theta\left[\Delta-d\left(\mathbf{s}_{i}\left(t\right),\mathbf{s}_{j}\left(t\right)\right)\right].\label{EQ6}
\end{align}
In Eq. (\ref{EQ6}), notion $\delta_{\cdot,\cdot}$ denotes the Kronecker delta function and $\Theta\left(\cdot\right)$ is the Heaviside step function. By simple calculation, we can transform Eq. (\ref{EQ2}) as 
\begin{align}
&\overline{\mathbf{s}}_{i}\left(t\right)-\mathbf{s}_{i}\left(t\right)\notag\\=&\frac{1}{\operatorname{deg}\left(\sigma_{i}\left(t\right)\right)+1}\left(\sum_{\sigma_{j}\in V}\mathbf{A}_{ij}\left(t\right)\mathbf{s}_{j}\left(t\right)-\operatorname{deg}\left(\sigma_{i}\left(t\right)\right)\mathbf{s}_{i}\left(t\right)\right),\label{EQ7}
\end{align}
where $\operatorname{deg}\left(\cdot\right)$ denotes the degree in a graph defined by the adjacency matrix in Eq. (\ref{EQ6}).

If we consider the observable global state 
\begin{align}
&\mathbf{s}\left(t\right)=\left[\mathbf{s}_{1}\left(t\right),\ldots,\mathbf{s}_{\vert V\vert}\left(t\right)\right]^{\mathsf{T}},\label{EQ8}
\end{align}
where $\mathsf{T}$ denotes the transpose, we can reformulate Eq. (\ref{EQ7}) into a new form
\begin{align}
&\overline{\mathbf{s}}_{i}\left(t\right)-\mathbf{s}_{i}\left(t\right)\notag\\=&\frac{1}{\operatorname{deg}\left(\sigma_{i}\left(t\right)\right)+1}\left(\left[\mathbf{A}_{i1}\left(t\right),\ldots,\mathbf{A}_{i\vert V\vert}\left(t\right)\right]-\mathbf{q}_{i}\left(t\right)\right)\mathbf{s}\left(t\right)\label{EQ9}
\end{align}
using vector $\mathbf{q}_{i}\left(t\right)$
\begin{align}
&\mathbf{q}_{i}\left(t\right)=\left[\delta_{i,1}\operatorname{deg}\left(\sigma_{1}\left(t\right)\right),\ldots,\delta_{i,\vert V\vert}\operatorname{deg}\left(\sigma_{\vert V\vert}\left(t\right)\right)\right].\label{EQ10}
\end{align}
After defining the observable of all local mean states
\begin{align}
&\overline{\mathbf{s}}\left(t\right)=\left[\overline{\mathbf{s}}_{1}\left(t\right),\ldots,\overline{\mathbf{s}}_{\vert V\vert}\left(t\right)\right]^{\mathsf{T}}\label{EQ11}
\end{align}
and the observable of the inverse of the degrees of all units
\begin{align}
&\mathbf{q}\left(t\right)=\operatorname{diag}\left(\left[\frac{1}{\operatorname{deg}\left(\sigma_{1}\left(t\right)\right)+1},\ldots,\frac{1}{\operatorname{deg}\left(\sigma_{\vert V\vert}\left(t\right)\right)+1}\right]\right)\label{EQ12}
\end{align}
using the diagonal $\operatorname{diag}\left(\cdot\right)$, we can eventually derive a global version of Eq. (\ref{EQ4})
\begin{align}
\overline{\mathbf{s}}\left(t\right)-\mathbf{s}\left(t\right)=&-\mathbf{q}\left(t\right)\mathbf{L}\left(t\right)\mathbf{s}\left(t\right),\label{EQ13}
\end{align}
where $\mathbf{L}\left(t\right)$ is the time-dependent Laplacian of the corresponding graph
\begin{align}
\mathbf{L}_{ij}\left(t\right)=\delta_{i,j}\sum_{k}\mathbf{A}_{ik}\left(t\right)-\mathbf{A}_{ij}\left(t\right).\label{EQ14}
\end{align}
Please see Fig. 1(b) for illustrations.

Now, let us consider the global dynamics characterized by Eq. (\ref{EQ13}). We define an abstract velocity vector to quantify the intrinsic degrees of the willingness of units to exhibit convergent evolution in the state space
\begin{align}
&\mathbf{v}_{c}\left(t\right)=\left[\mathbf{v}_{c,1}\left(t\right),\ldots,\mathbf{v}_{c,\vert V\vert}\left(t\right)\right]^{\mathsf{T}}.\label{EQ15}
\end{align}
Under the assumption of time continuity, the dynamics for $\mathbf{s}\left(t\right)$ to approach $\overline{\mathbf{s}}\left(t\right)$ during convergent evolution is formalized as
\begin{align}
\frac{\partial}{\partial t}\mathbf{s}\left(t\right)=-\mathbf{v}_{c}\left(t\right)\odot\mathbf{q}\left(t\right)\mathbf{L}\left(t\right)\mathbf{s}\left(t\right),\label{EQ16}
\end{align}
where local topology effects are described by $\mathbf{q}\left(t\right)$ (i.e., how the neighbors of a unit affect its evolution), and $\odot$ denotes the Schur product.

\paragraph{Global dynamics of divergent evolution.} Then we turn to analyze the divergent evolution processes of swarm behaviors. At the first glance, one only needs to consider the intrinsic degrees of the willingness of units to exhibit divergent evolution 
\begin{align}
&\mathbf{v}_{d}\left(t\right)=\left[\mathbf{v}_{d,1}\left(t\right),\ldots,\mathbf{v}_{d,\vert V\vert}\left(t\right)\right]^{\mathsf{T}}\label{EQ17}
\end{align}
and assume that units depart from associated time-dependent local mean states by moving in a direction opposite to Eq. (\ref{EQ16}). Under this condition, the joint dynamics of convergent and divergent evolution is nothing more but a Newtonian motion in the state space, $\frac{\partial}{\partial t}\mathbf{s}\left(t\right)=-\left[\mathbf{v}_{c}\left(t\right)-\mathbf{v}_{d}\left(t\right)\right]\odot\mathbf{q}\left(t\right)\mathbf{L}\left(t\right)\mathbf{s}\left(t\right)$. 

 \begin{figure*}[!t]
\includegraphics[width=1\columnwidth]{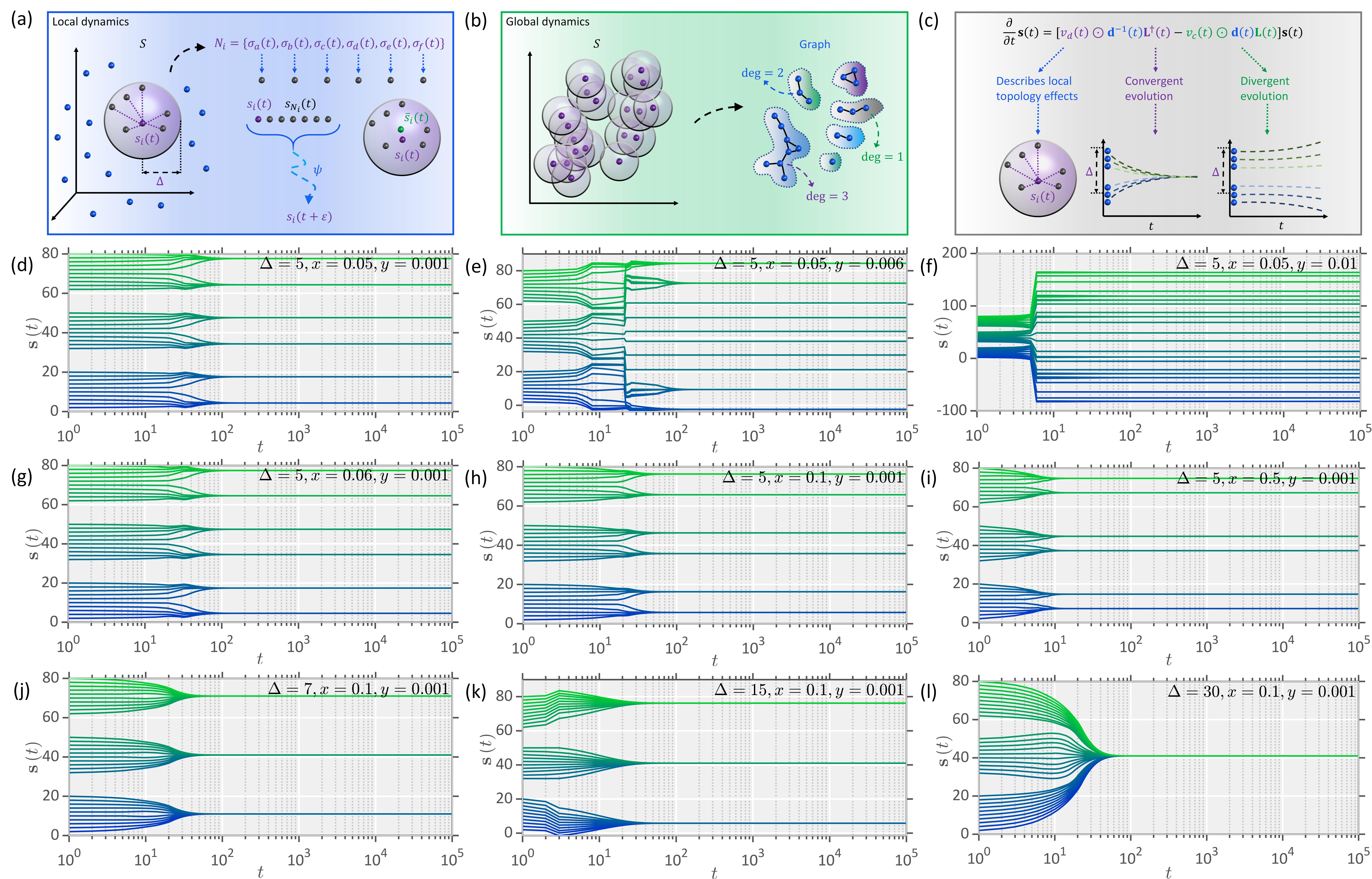}
\caption{Conceptual illustrations of the Laplacian dynamics of convergent and divergent swarm behaviors and the computational experiments with regular initialization. (a) On a local scale, the dynamics of each unit is determined by the interactions between it and its neighbors. (b) Based on the underlying graph structure, the local dynamics of every unit can be related to the global dynamics. (c) The joint dynamics on a global scale is shaped by local topology effects, the tendencies of convergent evolution, and the tendencies of divergent evolution. (d-l) The results of computational experiments with regular initialization and different parameter settings are visualized. A homogeneous system with uniform motions is defined on $30$ units, whose states are defined as $1$-dimensional to offer a clear vision. In the regular initialization, these units are divided equally into $3$ groups. The distance between units within each group is set as $2$ and the distance between adjacent groups is set as $10$. The experiments run $10^{5}$ iterations under each condition.} 
\end{figure*} 

However, it is unreasonable for our framework to be such trivial because Newtonian mechanics may be unpractical in abstract state space. The information of the mentioned opposite direction can be unavailable in real cases. For instance, if the state space describes the preference of drinks, it is impossible to precisely define an opposite direction of preferring coffee, i.e., it is difficult for units to tell preferring which drink is opposite to preferring coffee. This is because the definition of an opposite drink of coffee is ill-posed, at least in most situations. To describe these non-negligible cases where Newtonian mechanics is unacceptable, we need to consider a weaker assumption.

 \begin{figure*}[!t]
\includegraphics[width=1\columnwidth]{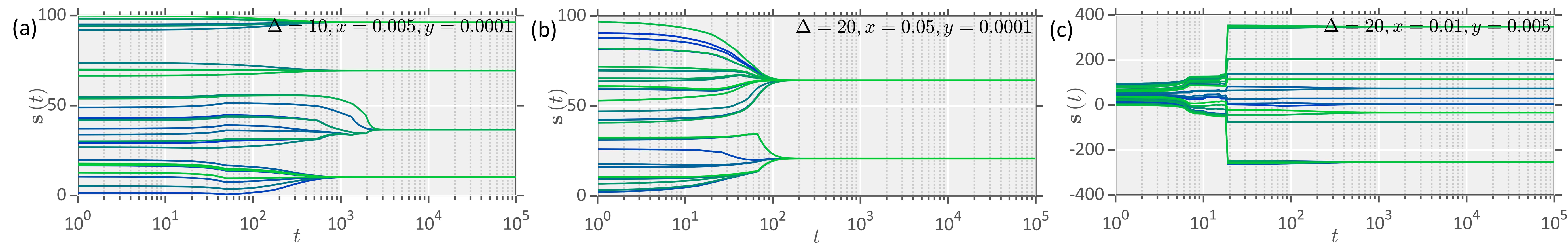}
\caption{The computational experiments with random initialization. (a-c) The results of the computational experiments are derived based on a homogeneous system of $30$ units with uniform motions in a $1$-dimensional state space. In the random initialization, each unit is assigned with an initial coordinate in the state space that is randomly selected following a uniform distribution in $\left[0,100\right]$. The experiments run $10^{5}$ iterations under each condition.} 
\end{figure*} 

\begin{figure*}[!t]
\includegraphics[width=1\columnwidth]{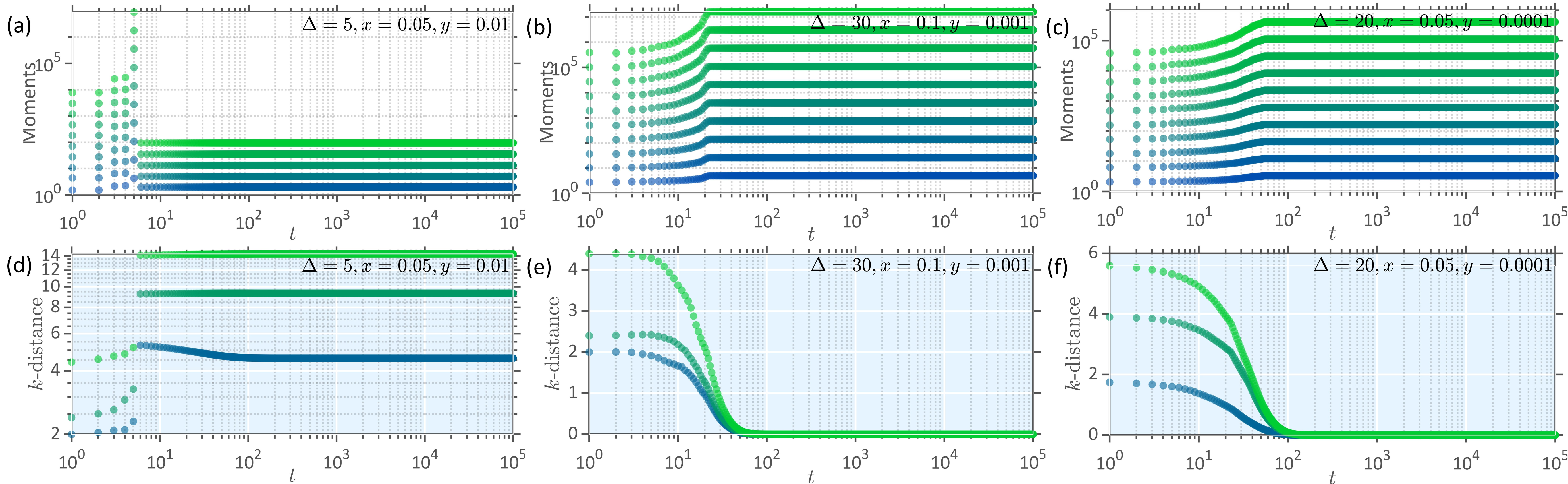}
\caption{The eigenvalue spectrum of matrix $\mathbf{H}\left(t\right)$ and the average $k$-distance in state space. (a-c) The observable of matrix $\mathbf{H}\left(t\right)$ is derived using the homogeneous systems generated in Fig. 1(f), Fig. 1(l), and Fig. 2(b), where we measure the $m$-th standardized moments of the eigenvalue spectrum. The measured results of $m\in\{3,\ldots,12\}$ are marked by colors, where the color changes from blue to green as $m$ increases. (d-f) The average $k$-distances in state space, where $k\in\{2,3,4\}$, are measured as the functions of time. The color changes from blue to green as $k$ increases.} 
\end{figure*} 

Our idea arises from the relation between $\mathbf{L}\left(t\right)$ and $-\mathbf{L}^{\dagger}\left(t\right)$, where $\dagger$ denotes the Moore–Penrose pseudoinverse \cite{barata2012moore}. To understand their relation, let us consider an instance where the covariance matrix units is defined as $\mathbf{C}\left(t\right)=\mathbf{L}\left(t\right)$. Two units are expected to evolve inversely (i.e., they have a negative covariance) if they are connected by an edge in the graph defined by Eq. (\ref{EQ6}). In this case, adjusting edges is equivalent to adjusting the differences (i.e., gradients) between units. Therefore, Eq. (\ref{EQ16}) can utilize $-\mathbf{L}\left(t\right)$ in the derivative to minimize those differences. If we further consider the associate precision matrix $\mathbf{Q}\left(t\right):=\mathbf{C}^{\dagger}\left(t\right)=\mathbf{L}^{\dagger}\left(t\right)$, we can relate the partial correlation \cite{rue2005gaussian,lawrance1976conditional} between units $\sigma_{i}$ and $\sigma_{j}$ with $\mathbf{L}^{\dagger}\left(t\right)$
\begin{align}
\operatorname{corr}\left(\sigma_{i},\sigma_{j}\vert V\setminus\{\sigma_{i},\sigma_{j}\}\right)=&-\frac{\mathbf{Q}_{ij}\left(t\right)}{\sqrt{\mathbf{Q}_{ii}\left(t\right)\mathbf{Q}_{jj}\left(t\right)}},\label{EQ18}\\=&-\frac{\mathbf{L}^{\dagger}_{ij}\left(t\right)}{\sqrt{\mathbf{L}^{\dagger}_{ii}\left(t\right)\mathbf{L}^{\dagger}_{jj}\left(t\right)}}.\label{EQ19}
\end{align}
Based on Eq. (\ref{EQ19}), two units are expected to evolve consistently (i.e., have a positive partial correlation) if they correspond to a negative value in $\mathbf{L}^{\dagger}\left(t\right)$ (i.e., a positive value in $-\mathbf{L}^{\dagger}\left(t\right)$). In this case, adjusting edges is equivalent to controlling similarities between units. During divergent evolution, units are expected to minimize their similarities, which can be realized by the term $\mathbf{L}^{\dagger}\left(t\right)$ in the derivative of the following dynamics
\begin{align}
\frac{\partial}{\partial t}\mathbf{s}\left(t\right)=\mathbf{v}_{d}\left(t\right)\odot\mathbf{q}^{-1}\left(t\right)\mathbf{L}^{\dagger}\left(t\right)\mathbf{s}\left(t\right).\label{EQ20}
\end{align}
In Eq. (\ref{EQ20}), we use $\mathbf{q}^{-1}\left(t\right)$, the inverse matrix of $\mathbf{q}\left(t\right)$, to characterize local topology effects.

\paragraph{Joint global dynamics of convergent and divergent evolution.} Now, we have the opportunity to define the joint global dynamics of convergent and divergent evolution based on Eq. (\ref{EQ16}) and Eq. (\ref{EQ20}). Specifically, the joint dynamics is given as
\begin{align}
\frac{\partial}{\partial t}\mathbf{s}\left(t\right)=&\big[\mathbf{v}_{d}\left(t\right)\odot\mathbf{q}^{-1}\left(t\right)\mathbf{L}^{\dagger}\left(t\right)\notag\\&-\mathbf{v}_{c}\left(t\right)\odot\mathbf{q}\left(t\right)\mathbf{L}\left(t\right)\big]\mathbf{s}\left(t\right),\label{EQ21}
\end{align}                                    
which is essentially a type of generalized Laplacian dynamics with more complicated details (see Refs. \cite{altafini2014predictable,fax2004information,veerman2020primer} for the elementary Laplacian dynamics). See Fig. 1(c) for illustrations. Although the idea underlying Eq. (\ref{EQ21}) is natural and simple, analyzing the dynamic properties of Eq. (\ref{EQ21}) is intricate because the competition between convergent and divergent evolution is characterized by the non-trivial interaction between $\mathbf{L}\left(t\right)$ and $\mathbf{L}^{\dagger}\left(t\right)$ under the effects of local topology. To offer a clear vision, we primarily consider a special case of Eq. (\ref{EQ21}) in our subsequent analysis. 

\begin{figure*}[!t]
\includegraphics[width=1\columnwidth]{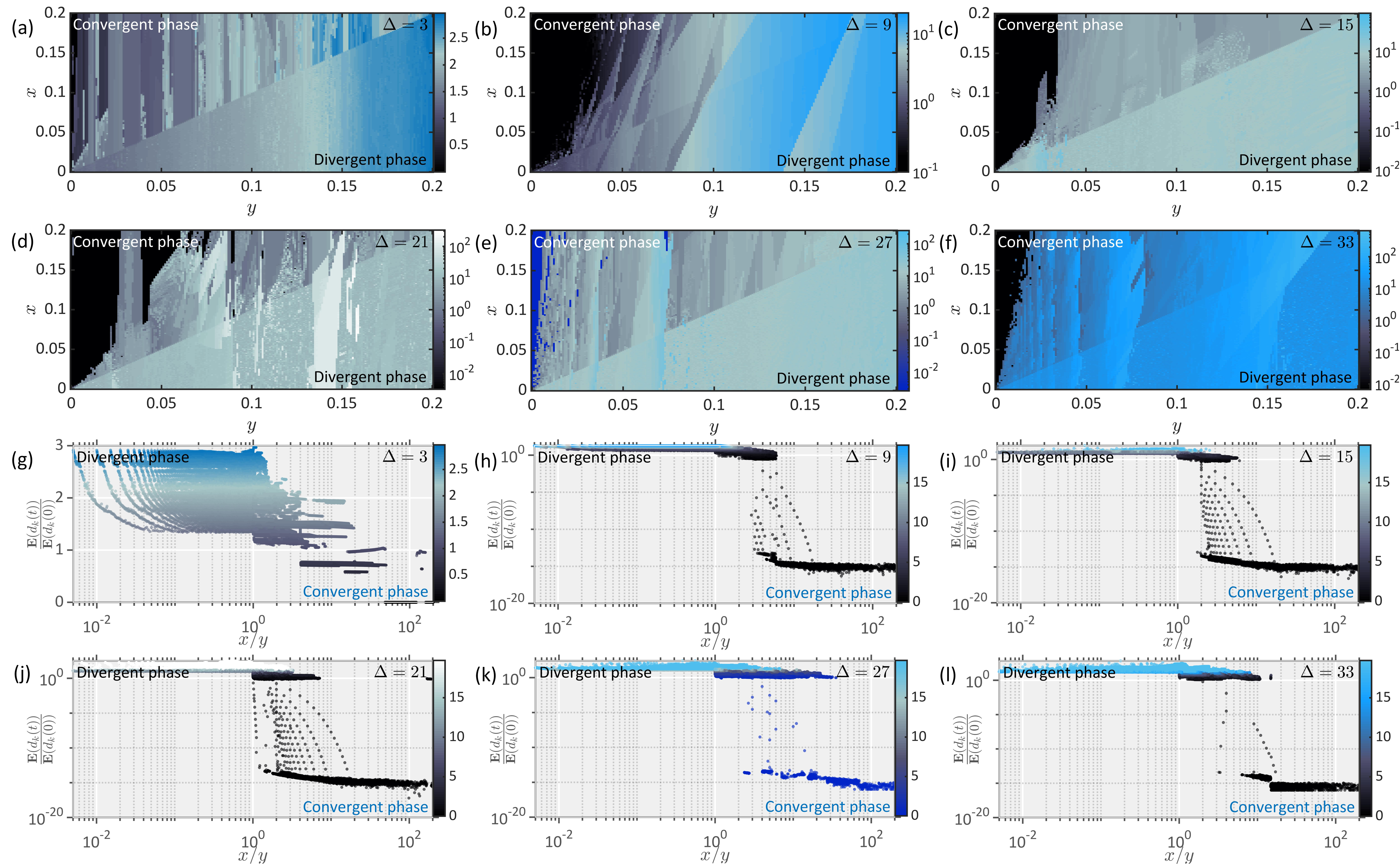}
\caption{The phase space of convergent and divergent evolution. (a-f) Given each condition of $\Delta\in\{3,9,15,21,27,33\}$, we generate a homogeneous system with uniform motions and regular initialization for every $\left(x,y\right)\in\left[0.001,0.2\right]^{2}$ and run the experiment for $10^{5}$ iterations. The generated results support the measurement of ratio $\frac{\mathbb{E}\left(d_{k}\left(t\right)\right)}{\mathbb{E}\left(d_{k}\left(0\right)\right)}$ ($k=3$). The heat maps of this ratio under all conditions are shown in (a-f), where different color maps are used to distinguish between different settings of $\Delta$. (g-l) Ratio $\frac{\mathbb{E}\left(d_{k}\left(t\right)\right)}{\mathbb{E}\left(d_{k}\left(0\right)\right)}$ is shown as a function of $\frac{x}{y}$ under each experiment condition, where color maps are consistent with (a-f).} 
\end{figure*} 

\paragraph{A special case on a homogeneous system with uniform motions.} The considered special case of Eq. (\ref{EQ21}) is a homogeneous system of units that exhibits uniform motions in the state space. To create uniform motions (i.e., with constant velocities) and make the system homogeneous (here, being homogeneous refers to the case where all units share the same intrinsic properties), we need to define 
\begin{align}
\mathbf{v}_{c}\left(t\right)\equiv x\mathbf{1}\mathbf{1}^{\mathsf{T}},\;\exists x\in\mathbb{R},\;\forall t\in T,\label{EQ22}\\
\mathbf{v}_{d}\left(t\right)\equiv y\mathbf{1}\mathbf{1}^{\mathsf{T}},\;\exists y\in\mathbb{R},\;\forall t\in T.\label{EQ23}
\end{align} 

Therefore, in a homogeneous system with uniform motions, Eq. (\ref{EQ21}) reduces to a simple form
\begin{align}
\frac{\partial}{\partial t}\mathbf{s}\left(t\right)&=\left[y\mathbf{q}^{-1}\left(t\right)\mathbf{L}^{\dagger}\left(t\right)-x\mathbf{q}\left(t\right)\mathbf{L}\left(t\right)\right]\mathbf{s}\left(t\right),\label{EQ24}
\\&:=\mathbf{H}\left(t\right)\mathbf{s}\left(t\right)\label{EQ25}
\end{align}   
which is easier to analyze. In computational experiments, we apply the following numerical iteration
\begin{align}
\mathbf{s}\left(t+\varepsilon\right)&=\varepsilon\mathbf{H}\left(t\right)\mathbf{s}\left(t\right)+\mathbf{s}\left(t\right)\label{EQ26}
\end{align} 
to derive the results of Eq. (\ref{EQ25}). Note that Eq. (\ref{EQ26}) only serves as a simple illustration. One can also consider more complicated numerical approaches. 

For the sake of visualization, we mainly consider a $1$-dimensional state space in our work, yet a state space with higher dimensions can be directly analyzed without any modification. In Figs. 1(d-l), we present diverse instances of this homogeneous system with different parameter settings, where a regular initialization condition of the system is applied to offer a clear vision (i.e., initialized units are uniformly distributed and form clear cluster structures in the state space). As shown in Figs. 1(d-i), different ratios between $x$ and $y$ in Eq. (\ref{EQ26}) create distinct trends of convergent and divergent swarm behaviors. A larger $x$ can lead to convergent evolution while a larger $y$ makes units divergent. As suggested by Figs. 1(j-l), the change of local interaction range controlled by a varying $\Delta$ in Eq. (\ref{EQ2}) leads to distinct final patterns in the state space after evolution. A larger local interaction range can amplify the trends of convergent and divergent evolution because more units are involved. These phenomena can also be observed in Fig. 2, where a random initialization condition is applied to the system (i.e., the initial states of units are randomly sampled following a specific distribution). These results provide qualitative illustrations of the global dynamics of swarm behaviors and suggest how the dynamics is shaped by the competition between convergent and divergent evolution trends or by the local interaction range.

\begin{figure*}[!t]
\includegraphics[width=1\columnwidth]{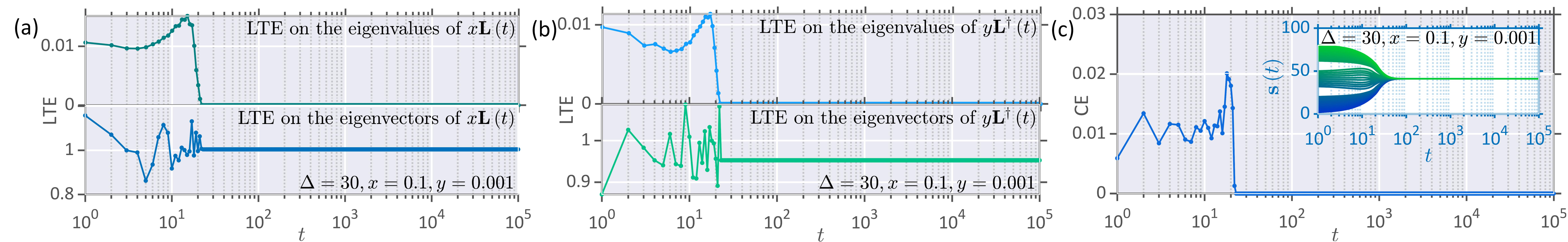}
\caption{The non-trivial properties of the Laplacian dynamics of convergent and divergent swarm behaviors. (a-b) The local topology effects (LTE) on the eigenvalues and eigenvectors of $x\mathbf{L}\left(t\right)$ and $y\mathbf{L}^{\dagger}\left(t\right)$ are shown as the functions of $t$ based on the homogeneous system generated in Fig. 1(l) (also see the inserted sub-plot in Fig. 5(c)). (c) The deviation of $\mathbf{H}\left(t\right)$ from being commutative is measured by the proposed commutator error (CE), where the applied homogeneous system is shown in the inserted sub-plot for reference.} 
\end{figure*} 

To quantitatively analyze the convergent and divergent evolution of swarm behaviors, we study the properties of the eigenvalue spectrum of $\mathbf{H}\left(t\right)$. Specifically, we analyze the standardized moments of the eigenvalue spectrum to reflect the latent dynamics of the evolution process (note that the first and the second standardized moment are not considered because they are constant \cite{larsen2005introduction}). Meanwhile, we consider $d_{k}\left(\mathbf{s}_{i}\left(t\right)\right)$, the distance between state $\mathbf{s}_{i}\left(t\right)$ and its $k$-nearest state at moment $t$ given a distance function $d\left(\cdot,\cdot\right)$ in Eq. (\ref{EQ2}). We define the average $k$-nearest distance at moment $t$ as
\begin{align}
    \mathbb{E}\left(d_{k}\left(t\right)\right)=\frac{1}{\vert V\vert}\sum_{\sigma_{i}\in V}d_{k}\left(\mathbf{s}_{i}\left(t\right)\right)\label{EQ27}
\end{align}
to characterize the global state of the system at moment $t$. In Fig. 3, these two kinds of metrics are calculated on three representative instances of the homogeneous system in Eq. (\ref{EQ25}). As shown in Fig. 3, distance $ \mathbb{E}\left(d_{k}\left(t\right)\right)$ exhibits distinct changes during convergent and divergent evolution (we define $d\left(\cdot,\cdot\right)$ as the Euclidean distance for convenience). Before the dynamics converges to its limit, the standardized moments of the eigenvalue spectrum dramatically increase during divergent evolution while decreasing during divergent evolution.

The results shown in Fig. 3 inspire us to distinguish between convergent and divergent phases using $\frac{\mathbb{E}\left(d_{k}\left(t\right)\right)}{\mathbb{E}\left(d_{k}\left(0\right)\right)}$, the relative ratio between the average $k$-nearest distances at moment $0$ and moment $t$. This ratio is larger than one in the divergent phase while it is smaller than one in the convergent phase. In Fig. 4, we analyze the phase space of the homogeneous system in terms of $\frac{\mathbb{E}\left(d_{k}\left(t\right)\right)}{\mathbb{E}\left(d_{k}\left(0\right)\right)}$ under different conditions of $x$, $y$, and $\Delta$. As shown in Figs. 4(a-f), convergent and divergent phases share a non-trivial and relatively blurry boundary on the plane of $\left(x,y\right)$, which changes across different local interaction range conditions. Phase-transition-like phenomena can be observed if we consider $\frac{\mathbb{E}\left(d_{k}\left(t\right)\right)}{\mathbb{E}\left(d_{k}\left(0\right)\right)}$ as an order parameter and treat $\frac{x}{y}$ as a control parameter in Figs. 4(g-l). As $\frac{x}{y}$ increases, a kind of transition from the divergent phase to the convergent phase occurs. However, it remains elusive whether these phenomena really arise from specific underlying phase transitions or from other unknown mechanisms. We cannot confirm the detailed properties of these potential phase transitions (e.g., the order of transition) yet. Below, we summarize the main difficulties for theoretically verifying the latent phase transition to suggest possible directions for future explorations.

\paragraph{Mathematical challenges for studying homogeneous system with uniform motions.} The main challenges for studying the latent phase transition arise from the non-triviality of predicting the limiting behaviors of the homogeneous system (i.e., the classification of convergent and divergent phases depends on the limit state of the system after the evolution process converges). 

Although the motions of units in the state space have been constrained as uniform, the homogeneous system in Eq. (\ref{EQ25}) is shaped by complicated local topology effects, which is in contrast to a more theoretically favorable case where local topology effects are completely excluded 
\begin{align}
\frac{\partial}{\partial t}\mathbf{s}\left(t\right)&=\left[y\mathbf{L}^{\dagger}\left(t\right)-x\mathbf{L}\left(t\right)\right]\mathbf{s}\left(t\right):=\mathbf{M}\left(t\right)\mathbf{s}\left(t\right).\label{EQ28}
\end{align} 
In such an ideal case, we have the opportunity to express $0,\omega_{2}\left(t\right)\leq\ldots\leq\omega_{\vert V\vert}\left(t\right)$, the eigenvalues of matrix $\mathbf{M}\left(t\right)$, based on the eigenvalues $0\leq \lambda_{2}\left(t\right)\leq\ldots\leq\lambda_{\vert V\vert}\left(t\right)$ of matrix $\mathbf{L}\left(t\right)$ 
\begin{align}
\omega_{i}\left(t\right)=\frac{y}{\lambda_{i}\left(t\right)}-x\lambda_{i}\left(t\right),\;\forall i\geq 2.\label{EQ29}
\end{align}  
Meanwhile, matrix $\mathbf{M}\left(t\right)$ shares the same group of eigenvectors, $\mathbf{1},\mathbf{u}_{2}\left(t\right),\ldots,\mathbf{u}_{\vert V\vert}\left(t\right)$, as matrix $\mathbf{L}\left(t\right)$. Eq. (\ref{EQ29}) is derived from the fact that the eigenvalues and eigenvectors of matrix $\mathbf{M}\left(t\right)$ are $0,\frac{1}{\lambda_{2}\left(t\right)},\ldots,\frac{1}{\lambda_{\vert V\vert}\left(t\right)}$ and $\mathbf{1},\mathbf{u}_{2}\left(t\right),\ldots,\mathbf{u}_{\vert V\vert}\left(t\right)$, respectively \cite{fontan2021properties,bullo2020lectures}. The explicit eigen-decomposition of $\mathbf{M}\left(t\right)$ provides rich information about the evolution dynamics, making it possible to analyze Eq. (\ref{EQ28}) following the dynamic system theory \cite{bullo2020lectures}. However, this possibility vanishes due to the local topology effects in Eq. (\ref{EQ25}). To quantify these non-trivial effects, we can measure the cosine distance between the vector of eigenvalues of $y\mathbf{L}^{\dagger}\left(t\right)$ and that of $y\mathbf{q}^{-1}\left(t\right)\mathbf{L}^{\dagger}\left(t\right)$ (similarly, between the vector of eigenvalues of $x\mathbf{L}\left(t\right)$ and that of $x\mathbf{q}\left(t\right)\mathbf{L}^{\dagger}\left(t\right)$). Meanwhile, we can calculate the average cosine distance between the eigenvectors of $y\mathbf{L}^{\dagger}\left(t\right)$ and those of $y\mathbf{q}^{-1}\left(t\right)\mathbf{L}^{\dagger}\left(t\right)$ (similarly, between the eigenvectors of $x\mathbf{L}\left(t\right)$ and those of $x\mathbf{q}\left(t\right)\mathbf{L}^{\dagger}\left(t\right)$). These cosine distances reflect the local topology effects on the eigenvalues and eigenvectors of $x\mathbf{L}\left(t\right)$ and $y\mathbf{L}^{\dagger}\left(t\right)$. In Figs. 5(a-b), we present an instance calculated on a representative homogeneous system, suggesting that these non-negligible local topology effects are volatile until the evolution dynamics converges. Therefore, Eq. (\ref{EQ29}) departs from the actual properties of Eq. (\ref{EQ25}) in most cases.

Moreover, another challenge arises from that $\mathbf{H}\left(t\right)$ in Eq. (\ref{EQ25}) is not commutative until the evolution dynamics converges to its limit. We can measure the deviation of $\mathbf{H}\left(t\right)$ from being commutative based on
\begin{align}
\chi\left(t,t^{\prime}\right)=\sum_{i,j}\vert\left[\mathbf{H}\left(t\right),\mathbf{H}\left(t^{\prime}\right)\right]_{ij}\vert,\label{EQ30}
\end{align} 
where $\left[\cdot,\cdot\right]$ denotes the commutator and $\left[\cdot,\cdot\right]_{ij}$ is the $\left(i,j\right)$-th element of the commutator. We refer to $\chi\left(\cdot,\cdot\right)$ as the commutator error. In Fig. 5(c), we calculate an instance of $\chi\left(\cdot,\cdot\right)$ by setting $t^{\prime}=t+\varepsilon$. The derived results suggest that $\mathbf{H}\left(t\right)$ becomes trivially commutative only after it arrives at its limit. The absence of a commutative $\mathbf{H}\left(t\right)$ in most cases makes it invalid to express the solution of Eq. (\ref{EQ25}) as $\mathbf{s}\left(t\right)=\exp\left(\int_{0}^{t}\mathbf{H}\left(\tau\right)\mathsf{d}\tau\right)\mathbf{s}\left(0\right)$, rejecting the possibility of simplifying Eq. (\ref{EQ25}).

In sum, the above properties create obstacles for theoretically analyzing the homogeneous system in Eq. (\ref{EQ25}). The situation becomes even more non-trivial if we relax the constraints on $\mathbf{v}_{c}$ and $\mathbf{v}_{d}$ to consider heterogeneous cases. These challenges may serve as valuable directions for future theoretical studies.

    \paragraph{Conclusion.} In this study, we propose a minimal model to describe the convergent and divergent evolution of swarm behaviors. Despite its mathematical simplicity, the model exhibits abundant non-trivial behaviors and suggests the potential for phase transitions or similar phenomena between convergent and divergent evolution controlled by local interactions. The non-triviality of the proposed model may inspire new mathematical techniques in analyzing the physics of swarming phenomena. Moreover, the simplicity of the proposed model in computational aspects enables it to serve as a foundation for developing large-scale numerical simulations of swarm behaviors. 

 \paragraph{Acknowledgements.} This project is supported by the Artificial and General Intelligence Research Program of Guo Qiang Research Institute at Tsinghua University (2020GQG1017) as well as the Tsinghua University Initiative Scientific Research Program. 


\bibliography{apssamp}
\end{document}